\documentclass[twoside]{ilcws10}
\usepackage[latin1]{inputenc}
\usepackage[dvips]{graphicx,epsfig,color}
\usepackage{wrapfig,rotating}
\usepackage{amssymb,amsmath,array}

\begin{document}
\title{
Chargino and Neutralino Masses at ILC} 
\author{Yiming Li and Andrei Nomerotski
\vspace{.3cm}\\
University of Oxford - Sub-department of Particle Physics \\
Denys Wilkinson Building, Keble Road, Oxford OX1 3RH - UK
}

\newcommand{\charginop} {\tilde{\chi} _1^+}
\newcommand{\charginom} {\tilde{\chi} _1^-}
\newcommand{\charginopm} {\tilde{\chi}_1^{\pm}}
\newcommand{\neutralinoone} {\tilde{\chi} _1^0}
\newcommand{\neutralinotwo} {\tilde{\chi} _2^0}

\maketitle

\begin{abstract}
The chargino/neutralino pair production is one of the benchmarking processes of ILC. These processes are interesting not only because it allows high precision measurement of chargino and neutralino masses, but also for the reason that the separation of W and Z bosons through their hadronic decay products requires excellent jet resolution being a good benchmark of the detector performance. The analysis based on the SiD detector concept with four jets and missing energy final state will be presented. The uncertainty of chargino and neutralino cross sections can be determined with precision of 0.9\% and 4.2\% respectively. The mass uncertainties are obtained with a template fitting method achieving precision of better than 1 GeV.
\end{abstract}

\section{Introduction}

\begin{wraptable}{r}{0.4\columnwidth}
\centerline{\begin{tabular}{cc}
\hline  parameter & value \\
\hline $m_0$ & 206 GeV \\
$m_{1/2}$ & 293 GeV \\
tan $\beta$ & 10 \\
A & 0 \\
$\mu$ & 375 GeV \\
$M_{\neutralinoone}$ & 115.7 GeV \\
$M_{\charginopm}$ & 216.5 GeV \\
$M_{\neutralinotwo}$ & 216.7 GeV \\
\hline
\end{tabular}}
\caption{SUSY point 5 parameters.}\label{tab:parameter}
\end{wraptable}
The chargino/neutralino pair production or so-called SUSY Point 5 is one of the benchmarking processes proposed for the ILC~\cite{ILCbenchmarks}. For the SUSY parameters listed in Table~\ref{tab:parameter}, the lightest chargino mass is smaller than slepton masses but at the same time is much larger than the lightest neutralino mass. $\neutralinoone$ is the Lightest Supersymmetry Particle (LSP), therefore charginos dominantly decay into LSP and on-shell W bosons. In this scenario the $\neutralinotwo$ mass is almost degenerate with the chargino mass and it similarly decays into LSP and Z bosons.
\begin{eqnarray*}
e^+ e^-\;&\rightarrow&\;\charginop \charginom \;\rightarrow\;
\neutralinoone \neutralinoone W^+ W^- \\
e^+ e^-\;&\rightarrow&\;\neutralinotwo \neutralinotwo
\;\rightarrow\; \neutralinoone \neutralinoone Z^0 Z^0
\end{eqnarray*}

If we consider all-hadronic decays of the gauge bosons in the final state, the above processes will both be defined by a signature of four jets with large missing energy. Once the W or Z bosons are successfully reconstructed, their energy distribution will provide information on the chargino and neutralino masses. The sensitivity of the mass measurement is determined by template fitting. Since this measurement requires separation of pairs of W and Z bosons reconstructed from two jets, fine jet resolution and a good performance of Particle Flow Algorithm (PFA)~\cite{PFA} are essential.

A Monte-Carlo sample with $\sqrt{s}$ of 500 GeV and integrated luminosity of 500 fb$^{-1}$ was used for this study. It consisted of the chargino/neutralino signals and SUSY backgrounds such as $e^+e^-\rightarrow \neutralinoone\neutralinotwo$ and slepton pair production. Apart from this sample, a few more samples with the same statistics were generated in order to estimate the uncertainty of the mass measurement, each with one SUSY particle mass shifted by certain amount (Table~\ref{tab:susysample}). A Standard Model background sample at $\sqrt{s} = 500 $GeV was generated separately. All the samples have the same polarization, namely right handed 80\% electrons and left handed 30\% positions. The cross-section of chargino production with all hadronic final state is $57.1 fb$ and the neutralino one is $11.0 fb$. Note that the cross-section of neutralino pair production is much smaller than the chargino one, therefore the separation of the two, especially the neutralino selection in the presence of a large chargino background is challenging.
\begin{table}[h]
\centering
\begin{tabular}{cccc}
\hline Template & $M_{\neutralinoone}$ (GeV) & $M_{\charginopm}$ (GeV)& $M_{\neutralinotwo}$ (GeV)\\
\hline standard & 115.7 & 216.5 & 216.7 \\
$\neutralinoone + \Delta M$ & 115.7 + \textcolor{blue}{$\Delta M$} & 216.5 & 216.7\\
$\charginopm + \Delta M$ & 115.7 & 216.5 + \textcolor{blue}{$\Delta M$}& 216.7 \\
$\neutralinotwo + \Delta M$  & 115.7 & 216.5 & 216.7 + \textcolor{blue}{$\Delta M$}\\
\hline
\end{tabular}
\caption{MC sample parameters. $\textcolor{blue}{\Delta M} = -0.5 / 0.5 / 2 \, GeV$ or $6 \, GeV$ (only for $\neutralinotwo$)} \label{tab:susysample}
\end{table}

\section{Analysis Methods}
All samples used in this study were generated using the Monte Carlo program Whizard 1.4~\cite{Whizard}. PYTHIA 6.4~\cite{PYTHIA} is used for parton showering, fragmentation and decay to provide final state observable particles. The detector response to the generated events is simulated by the Linear Collider specific code SLIC~\cite{SLIC} based on the Geant4 toolkit~\cite{Geant4-1}~\cite{Geant4-2}. The full detector description of the baseline Silicon Detector (sid02~\cite{sid02}) is employed.

The particle reconstruction was performed within the lcsim 1.5 framework~\cite{lcsim}. Pattern recognition on hadronic final states used a PFA algorithm devised for SiD~\cite{SiDPFA}. The jet clustering and further physics analysis were performed within the MarlinReco framework~\cite{MarlinReco}. The Durham algorithm~\cite{Durham} was used for jet finding and the jet number was forced to be four rather than setting a $y_{cut}$ value.

The analysis strategy is defined as follows. Common preselections for the $\charginop \charginom$ and $\neutralinotwo \neutralinotwo$ events were defined to suppress the Standard Model background, as described in Section \ref{sec:preselection}. The cross-sections of both signal events could be determined after they were separated using the dijet mass correlation as described in Section \ref{sec:sig-separation}. The energy distributions of the gauge bosons as decay products gave information on the SUSY particle masses. Uncertainties of the masses from this measurement were determined by the template fitting method presented in Section \ref{sec:template-fitting}.

\subsection{Signal Preselection}
\label{sec:preselection}
The analysis started with selection of all-hadronic channel
events. All reconstructed particles were clustered into four jets. A
series of selections were applied (Table~\ref{tab:cuts}). The selection criteria are explained briefly as following:
\begin{itemize}
\item{Total visible energy: Events with missing energy are selected;}
\item{Number of tracks: Leptons reconstructed as jets with low multiplicity are eliminated;}
\item{Thrust; $cos\theta_{thrust}$: Jets originated from signal events should be distributed more uniformly;}
\item{$E_{jet}$; fraction of EM energy in each jet; lepton energy in each jet: Ruling out jets falsely reconstructed from energetic leptons;}
\item{$\theta(i,j)$: opening angle between jet $i$ and $j$ (ordered by energy);}
\item{Acoplanarity: The WW or ZZ pair from $\charginop\charginom$ or $\neutralinotwo\neutralinotwo$ decay are acoplanar, different from WW pairs in SM.}
\end{itemize}

\begin{table}[h]
\centering
\begin{tabular}{ll}
\hline cut & value  \\
\hline Jet number & = 4 \\
Total visible energy & $<$ 250 GeV \\
Number of tracks & $>$ 20\\
Thrust & $<$ 0.85\\
$\cos\theta_{thrust}$ & $<$ 0.9\\
$E_{jet}$ & $<$ 10 GeV \\
Fraction of EM energy in each jet & $<$ 80\% \\
$\theta$(1, 2) & $>$ 60$^{\circ}$\\
$\theta$(1, 3), $\theta$(1, 4) , $\theta$(1, 3)& $>$ 40$^{\circ}$\\
$\theta$(2, 4), $\theta$(3, 4) & $>$ 20$^{\circ}$\\
lepton energy in jet 1 & $<$ 40 GeV \\
lepton energy in jet 2 & $<$ 40 GeV \\
lepton energy in jet 3 & $<$ 30 GeV \\
lepton energy in jet 4 & $<$ 20 GeV \\
Acoplanarity of two reconstructed gauge bosons & $>$
$10^{\circ}$\\
\hline
\end{tabular}
\caption{Pre-selections. $\theta (i, j)$ defines the opening angle
between the $i$th and $j$th jet. Jets were ordered in energy,
e.g. jet 1 is the most energetic jet.} \label{tab:cuts}
\end{table}

\begin{figure}[htbp]
\centerline{\includegraphics[width=0.9\textwidth]{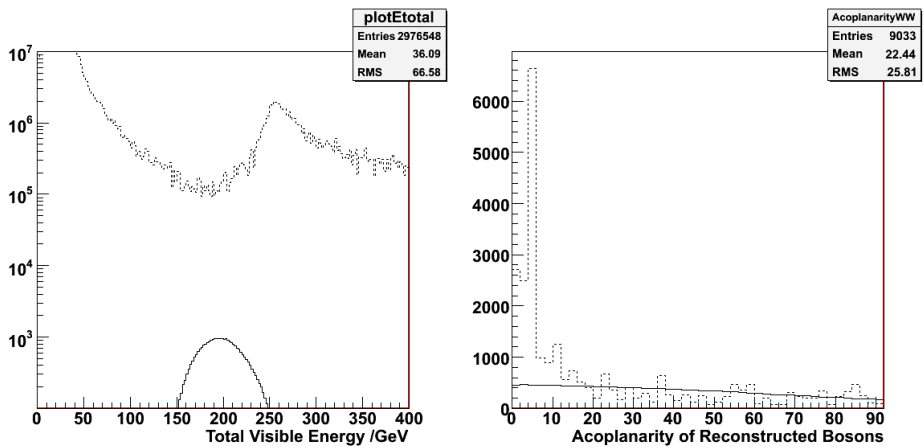}}
\caption{Examples of SM background suppression cuts for the chargino selection. Solid lines are the chargino signal and dashed lines are the Standard Model background. Left: Total visible energy; Right: Acoplanarity of two reconstructed W bosons.}
\label{fig:cuts1}
\end{figure}

As a result the Standard Model background was well suppressed (Figure~\ref{fig:cuts1}).

\subsection{Chargino/Neutralino Signal Separation}
\label{sec:sig-separation}
\begin{figure}[htbp]
\centerline{\includegraphics[width=0.9\textwidth]{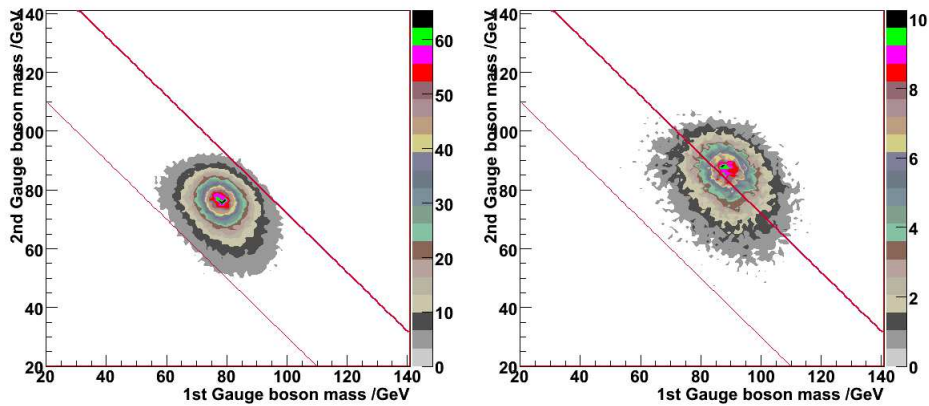}}
\caption{Reconstructed boson masses from the four jets, for chargino selection. Left: pure chargino signal; Right: pure neutralino signal. The region between the two straight lines shows an allowed chargino selection window.} \label{fig:mass2dCh}
\end{figure}

To reconstruct two gauge bosons using the four jets we have to
determine the jet pairing corresponding to correct combinations.
It is done by minimizing the quantity of $(m(j_1,
j_2)-m_{W/Z})^2+(m(j_3, j_4)-m_{W/Z})^2$, where $m(j_1, j_2)$ is the
invariant mass of the jets $j_1$ and $j_2$. All jets were ordered by energy and all possible jet permutations were considered. The W mass was used when selecting
$\charginop \charginom$ events, and the Z mass for $\neutralinotwo
\neutralinotwo$ events. The correlation of the two reconstructed
boson masses is shown in Figure~\ref{fig:mass2dCh}. Reconstructed boson masses populate different regions in the histogram for chargino and neutralino events, and their separation
can be achieved by a selection indicated by a thick line in
Figure~\ref{fig:mass2dCh}. Events in the lower left part ($m_{W1} + m_{W2} \leq 172 \, GeV$) were classified as $\charginop \charginom$ events and in the upper right part ($m_{W1} + m_{W2} > 172 \, GeV$) as $\neutralinotwo \neutralinotwo$ events. For chargino events
selection we also add a selection to remove events in the most lower
left corner of 2D mass histogram ($m_{W1} + m_{W2} > 130 \, GeV$), which are most likely due to the remaining leptonic or semi-leptonic decays of the bosons.

\subsection{Kinematic Fitting}

\begin{figure}
\centerline{\includegraphics[width=0.9\textwidth]{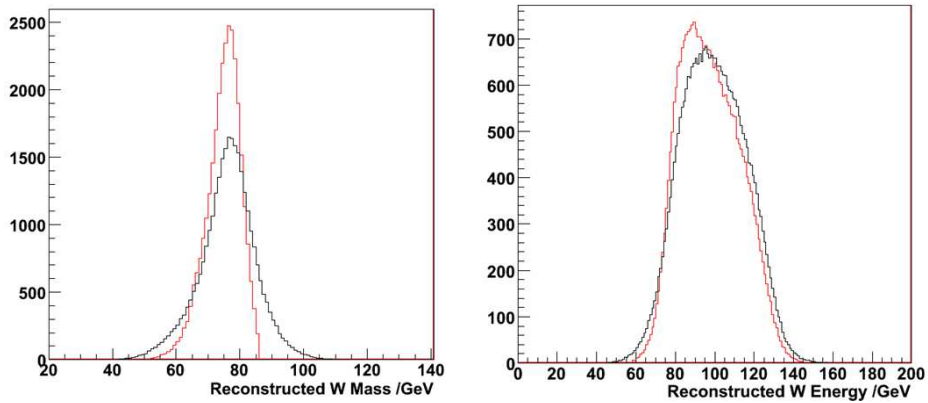}}
\caption{Mass (left) and energy (right) distributions of reconstructed W boson before (black) and after (red) kinematic fitting. }
\label{fig:kinfit}
\end{figure}

In the rest frame of $\charginopm$/$\neutralinotwo$, W$^{\pm}$/Z
bosons are monochromatic and their energy is determined by three masses:
$m_{\charginopm}/m_{\neutralinotwo}$, $m_{W^{\pm}}/m_Z$ and
$m_{\neutralinoone}$. For a boosted parent
$\charginopm$/$\neutralinotwo$, the boson energy distribution
depends on the SUSY particle masses which will determine the edges
of the energy spectrum. Our approach to obtain these masses is to
compare the energy spectrum with a reference MC sample template
(Table~\ref{tab:susysample}). Therefore the best possible energy
resolution is important for a high-precision mass measurement. Using
the fact that the two bosons reconstructed from four jets have
the same mass in both $\charginop\charginom$ and
$\neutralinotwo\neutralinotwo$ cases, kinematic fitting with one
constraint ($m_{boson1}=m_{boson2}$) can be used for improving the
energy distribution assuming that the detector resolution is known.
Kinfit in Marlinreco package~\cite{Kinfit} is used for the fitting. The comparison
of reconstructed W mass with and without fitting is shown in
Figure~\ref{fig:kinfit}.

\section{Cross-section Results}
\label{sec:cross-section}
\begin{figure}[htbp]
\centerline{\includegraphics[width=0.9\textwidth]{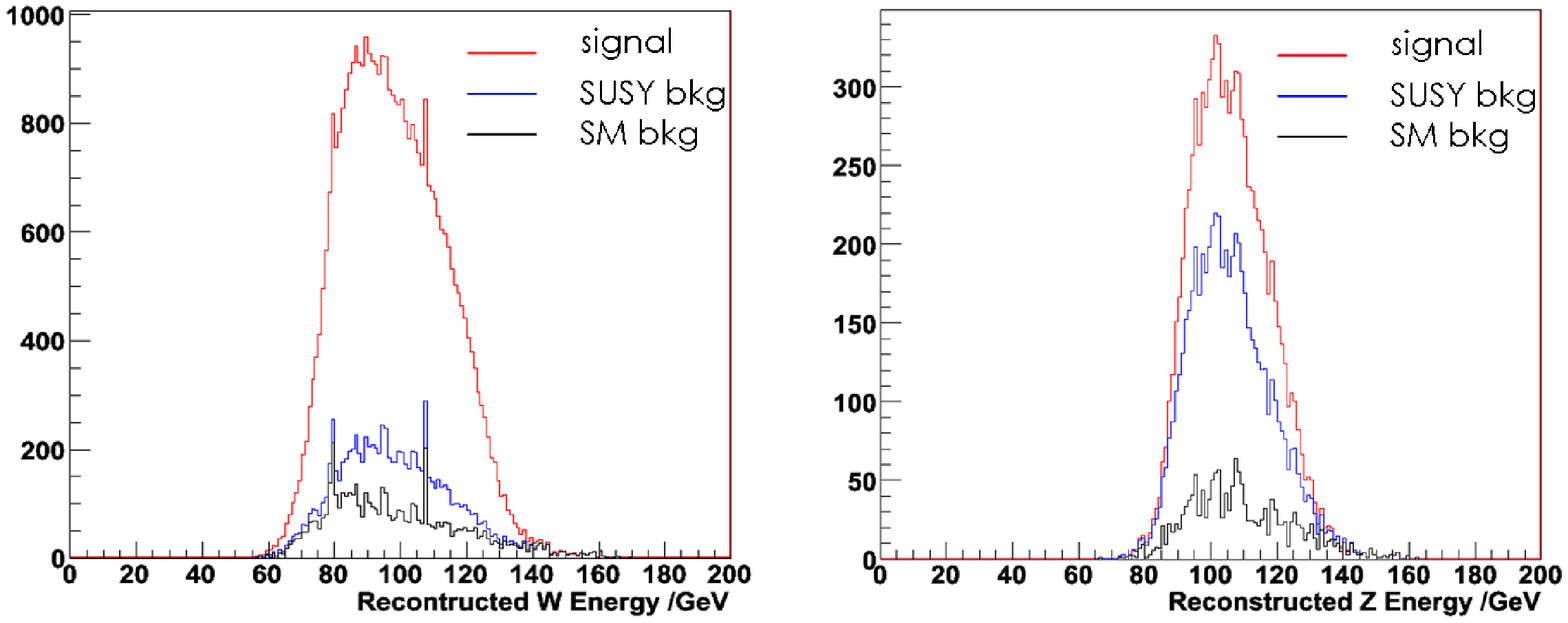}}
\caption{The inclusive histogram of reconstructed W energy for
$\charginop \charginom$ selection (left) and Z energy for $\neutralinotwo \neutralinotwo$ selection (right).}
\label{fig:selChNeu}
\end{figure}

The reconstructed boson energy distribution are shown in Figure~\ref{fig:selChNeu} after preselections and signal separtion. The chargino sample contains 15362 signal events, 1997 events of neutralino pair background, 65 events of other SUSY background and 2980 events of Standard Model background. Similarly the neutralino sample contains 1659 events of pure
neutralino signal, 2395 events of chargino background, 9 events of other SUSY background
and 865 events of SM background. The uncertainty of the cross section measurement
for chargino and neutralino pair production with all hadronic final state is 0.9\% and 4.2\% respectively neglecting the uncertainty of efficiencies.

\section{Mass Uncertainty and Template Fitting}
\label{sec:template-fitting}
\begin{figure}[htbp]
\centerline{\includegraphics[width=\textwidth]{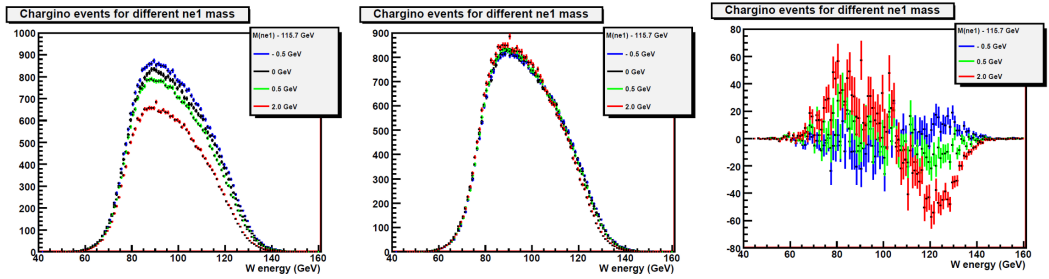}}
\caption{Left: W energy distribution for different $\neutralinoone$ masses; Middle: W energy distribution normalized to the same cross-section; Right: Difference of the normalized W energy spectrum with respect to the standard template ($M_{\neutralinoone} = 115.7 GeV$).}
\label{fig:W-varying-ne1}
\end{figure}

The determination of mass sensitivity can be explained on the example of the $\charginop\charginom$ process. The W energy distribution is shown in Figure \ref{fig:W-varying-ne1} for different $\neutralinoone$ masses. The were all normalized to the same cross section to explore the pure kinematic effect. The change of upper and lower edges of the spectrum after normalization is clearly visible. Assuming that we can generate a lot of templates, the mass of the SUSY particle masses can be obtained by comparing the boson energy spectrum from the data with those of various templates.

\begin{wrapfigure}{r}{0.3\columnwidth}
\centerline{\includegraphics[width=0.3\columnwidth]{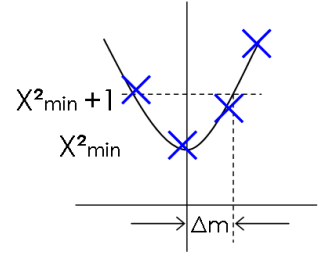}}
\caption{Sketch of template fitting method (not to scale).}
\label{Fig:template-fitting-sketch}
\end{wrapfigure}
In this analysis, one 'data' sample is taken from the standard template (Table \ref{tab:parameter}) without any SM background. The $\chi^2$ of a template with respect to the 'data' is defined as following:
\begin{equation}
\chi^2 = \sum^{Nbins}_{i=0} \frac{(y_{template,i}-y_{data,i}+\delta)^2}{\sigma^2_{template,i}+\sigma^2_{data,i}+\sigma^2_{SM,i}}
\end{equation}
where $y_{...i}$ denotes the content of the $i$th bin in the boson energy histogram. The $\delta$ term is added as a Gaussian smearing of the central value of $y_{template,i}-y_{data,i}$ to take the SM background into account. The $\chi^2$ dependence on the mass is parabolic, with the minimum located at the actual mass. The half-width of the $\chi^2 = \chi^2_{min} + 1$ gives one $\sigma$ uncertainty on the mass~\cite{LouisLyonsBook} as illustrated in Figure \ref{Fig:template-fitting-sketch}. The statistics of the templates are much larger ($\sim$ 10 times) than the 'data'. The uncertainty results are stable for different binning of the histogram, and the minimum $\chi^2_{min}$ over degree of freedom should be close to one.

\begin{figure}[htbp]
\centerline{\includegraphics[width=0.9\textwidth]{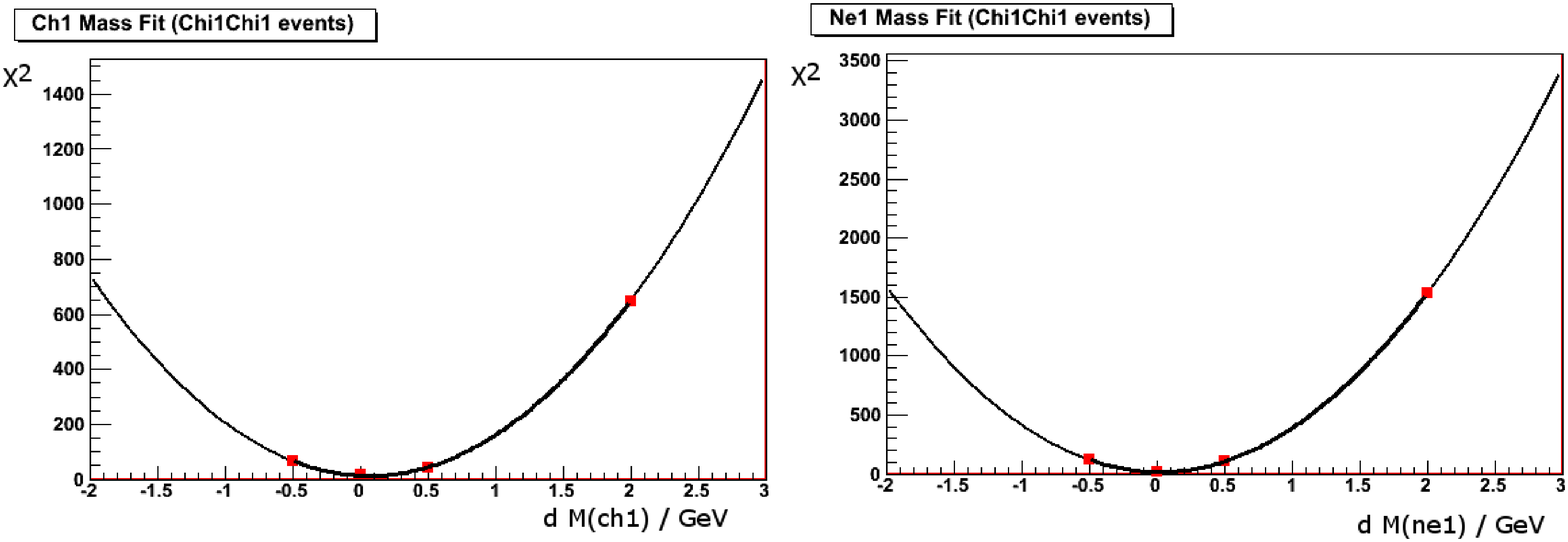}}
\caption{The $\chi^2$ dependence on the shifted mass of $\charginopm$ (left) and $\neutralinoone$ (right), calculated for the $\charginop \charginom$ signal. }
\label{fig:W-parabola}
\vspace{3mm}
\centerline{\includegraphics[width=0.9\textwidth]{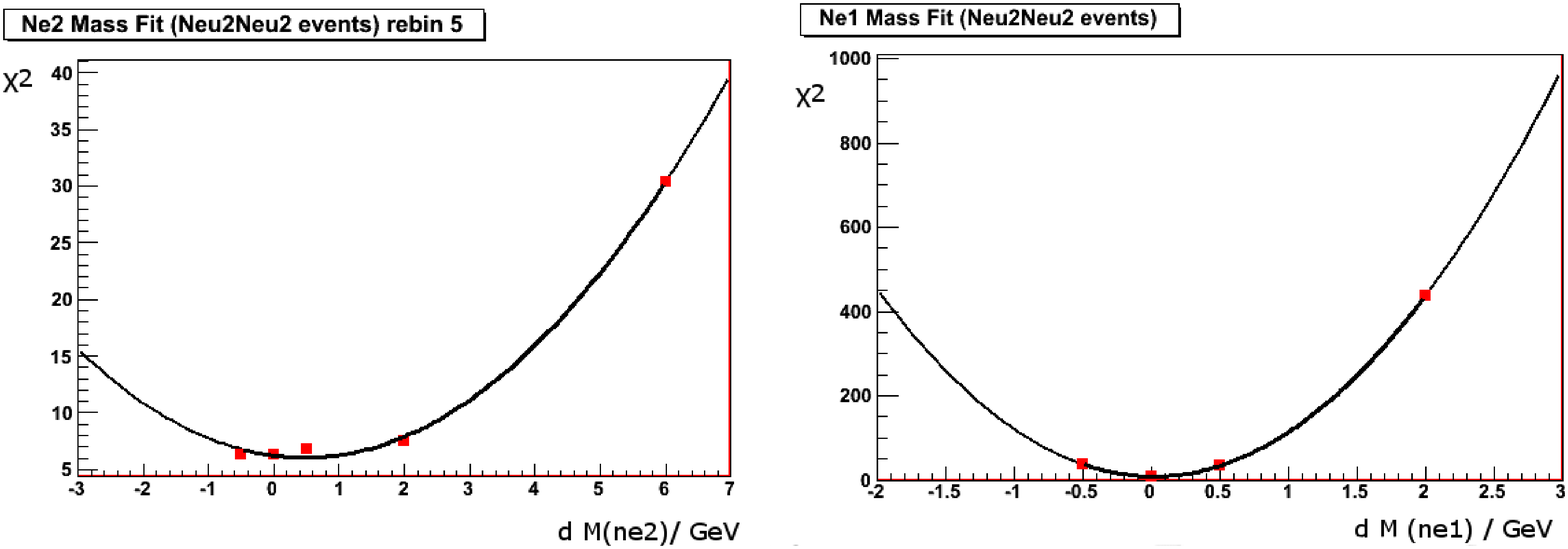}}
\caption{The $\chi^2$ dependence on the shifted mass of $\neutralinotwo$ (left) and $\neutralinoone$ (right), calculated for the $\neutralinotwo \neutralinotwo$ signal.}
\label{fig:Z-parabola}
\end{figure}

Figure \ref{fig:W-parabola} shows the $\chi^2$ parabola from the W energy spectrum, varying $\charginopm$ and $\neutralinoone$ masses respectively. It should be noticed that the point for $\Delta M = 0$ is calculated for an independent standard template other than the 'data' sample. Similarly for the $\neutralinotwo \neutralinotwo$ events the $\chi^2$ parabola can be obtained from various $\neutralinotwo$ and $\neutralinoone$ masses (Figure \ref{fig:Z-parabola}). $\Delta M$ has the the value of -0.5, 0, 0.5 and 2 GeV for all target masses, and 0.6 GeV only for $\neutralinotwo$. This is because the $\neutralinotwo \neutralinotwo$ sample has a worse purity comparing to $\charginop \charginom$ due to a large background of the latter, and therefore the change in Z energy edge is less sensitive to the $\neutralinotwo$ mass. These figures are obtained with the energy bin width of 5 GeV.

Uncertainties of SUSY particle masses can be estimated as in Table \ref{tab:mass-error} from the parabolas above. These are taken after averaging from different binning, namely energy histograms with 1 GeV, 2 GeV, 5 GeV or 10 GeV per bin.
\begin{table}[h]
\centering
\begin{tabular}{|cc|cc|}
\hline
$\charginop \charginom$ & & $\neutralinotwo \neutralinotwo$ & \\
\hline
$\Delta M( \charginopm )$ & 472 MeV & $\Delta M( \neutralinotwo )$ & 975 GeV \\
$\Delta M( \neutralinoone )$ & 156 MeV & $\Delta M( \neutralinoone )$ & 279 MeV \\
\hline
\end{tabular}
\caption{Mass uncertainty results.}
\label{tab:mass-error}
\end{table}

Each uncertainty was calculated by varying that specific mass alone, assuming other SUSY particle masses are measured perfectly. In order to estimate possible correlations between the mass uncertainties, several templates are needed with masses, e.g. $\charginopm / \neutralinoone$ or $\neutralinotwo / \neutralinoone$ shifted simultaneously. More samples will be generated for further study.

\section{Conclusion}
The chargino/neutralino pair production at the ILC with full SiD detector simulation are studied. The cross-section of $\charginop \charginom$ and $\neutralinotwo \neutralinotwo$ are measured with sensitivity to 0.9\% and 4.2\%. Template fitting method is applied to determine the uncertainty of the chargino and neutralino masses, and the mass uncertainties obtained were all below 1 GeV.


\begin{footnotesize}


\end{footnotesize}


\end{document}